\documentclass[aip,amsmath,amssymb,endfloats,
preprint
]{revtex4-1}
\usepackage{booktabs}
\usepackage{graphicx}
\usepackage[table,xcdraw]{xcolor}
\usepackage{subcaption}
\usepackage{chngcntr}

\usepackage{color}

\graphicspath{{./pre-print-image/}}
\def\hl#1{{#1}}

\begin{document}
\title{Deep learning-based design of broadband GHz complex and \hl{random} metasurfaces}
\author{Tianning Zhang}
\affiliation{Science, Mathematics and Technology, Singapore University of Technology and Design, 8 Somapah Road, Singapore 487372}
\author{Chun Yun Kee}
\affiliation{Science, Mathematics and Technology, Singapore University of Technology and Design, 8 Somapah Road, Singapore 487372}
\author{Yee Sin Ang}
\thanks{Authors to whom correspondence should be addressed: yeesin\_ang@sutd.edu.sg and ricky\_ang@sutd.edu.sg}
\affiliation{Science, Mathematics and Technology, Singapore University of Technology and Design, 8 Somapah Road, Singapore 487372}
\author{L. K. Ang}
\thanks{Authors to whom correspondence should be addressed: yeesin\_ang@sutd.edu.sg and ricky\_ang@sutd.edu.sg}
\affiliation{Science, Mathematics and Technology, Singapore University of Technology and Design, 8 Somapah Road, Singapore 487372}

\begin{abstract}
\hl{We are interested to explore the limit in using deep learning (DL) to study the electromagnetic response for complex and random metasurfaces, without any specific applications in mind. 
For simplicity, we focus on a simple pure reflection problem of a broadband electromagnetic (EM) plane wave incident normally on such complex metasurfaces in the frequency regime of 2 to 12 GHz. 
In doing so, we create a deep learning (DL) based framework called metasurface design deep convolutional neural network (MSDCNN) for both the forward and inverse design of three different classes of complex metasurfaces: (a) Arbitrary connecting polygons, (b) Basic pattern combination, and (c) Fully random binary patterns.}
The performance of each metasurface is evaluated and cross-benchmarked.
Dependent on the type of complex metasurfaces, sample size, and DL algorithms used, MSDCNN is able to provide good agreements and can be a faster design tool for complex metasurfaces as compared to the traditional full-wave electromagnetic simulation methods.
\hl{However, no single universal deep convolutional neural network (DCNN) model can work well for all metasurface classes based on detailed statistical analysis (such as mean, variance, kurtosis, mean squared error).} 
Our findings report important information on the advantages and limitation of current DL models in designing these ultimately complex metasurfaces.

\end{abstract}

\maketitle

\section{Introduction}
Metasurfaces are two-dimensional (2D) artificial structures that are designed and fabricated to manipulate the propagation of electromagnetic (EM) waves in order to have unique performance beyond the conventional materials \cite{quevedo2019roadmap,li2020metamaterials,li2021metasurfaces,cui2017information,khatib2021deep,qiu2019deep}.
It has attracted enormous research attention due to its extraordinary ability to control many electromagnetic properties, such as amplitude \cite{cheng2015structural,proust2016all}, phase\cite{khorasaninejad2017metalenses,zhang2021electrically,abdollahramezani2021electrically}, polarization\cite{zhang2019machine} and many
Various types of metasurfaces have been proposed, offering a large variety of specialized metasurfaces for different applications such as programmable metasurfaces \cite{bao2021programmable,zhang2020polarization,zhang2020optically}, transforming heat \cite{li2021transforming}, cloaking \cite{yang2016full,qian2021perspective}, hologram \cite{liu2020work}, conversion \cite{liu2016fully}, absorption \cite{mitrofanov2018efficient,mitrofanov2020perfectly}, scattering reduction \cite{zhang2021hyperuniform}, polarization \cite{khan2017ultra}, transmission \cite{you2020broadband} and others.

There are two general approaches in designing metasurfaces.
The first approach is the forward design, which is an iterative process involving parametric studies to explore within a given set of input parameters in order to produce the desired EM response or output.
A simulation tool (or forward numerical solver), which solves the underlying governing equations to provide reliable characterization of input parameters to match the calculated outputs.
The cost is determined largely by the simulation time of each trial and error.
If the number of input parameters is huge or to avoid computational cost, a designer often has to give up exhaustive exploration of the design space and settle on some trade-offs on the desired output.

The second approach is the inverse design, which is to find an optimal set of input parameters for a given output.
This is more difficult than the forward design as there is no definite or
unique solution for such a problem.
Thus, inverse design is typically formulated to search for the most approximate input conditions within a prescribed domain via an optimization algorithm.
Almost all of the inverse design problems are challenging, which require advanced algorithms, \hl{such as the heuristic algorithm of ant colony algorithm \cite{zhu2019optimal}, genetic algorithm \cite{jafar2018adaptive}, particle swarm algorithm \cite{zhang2017shaping} and topological optimization \cite{sui2015topology,zhu2020review,yang2017topology,phan2019high}}.

Machine learning (ML) techniques like deep learning (DL) \cite{lecun2015deep} has been successful in various fields involving complexity, such as computer vision, natural language processing, and speech signal processing.
Their applications in some traditional scientific disciplines have also grown significantly in recent years, including condensed matter physics \cite{carrasquilla2017machine}, particle physics \cite{baldi2014searching} , chemistry \cite{segler2018planning},
text mining for materials discovery \cite{nature2019}, discovering physical concepts \cite{iten2020discovering} and many other physics-based problems \cite{chen2016deep,zhou2017pdeep,gessulat2019prosit,PINN2019}.
DL based approaches for the design of metasurfaces are also gaining a lot of attentions \cite{khatib2021deep,qiu2019deep}, where various types of metasurfaces (with some prescribed regular patterns) have been successful designed \cite{liu2018training,an2020freeform,zhen2021realizing,deng2021neural,cardin2020surface,nadell2019deep, malkiel2018plasmonic,peurifoy2018nanophotonic,an2019deep,tahersima2019deep,asano2018optimization,jiang2019free,jiang2019global,haninverse,sajedian2019finding,liu2018generative,ma2018deep,zhu2021phase}.
Most of them are using techniques such as fully connected network (FCN), convolutional neural network (CNN), and \hl{transposed convolutional neural networks (t-CNN)}.
The FCN is composed of a series linear dense layers, and it is the most basic neural network, although the input and output of this network are limited to one-dimensional (1D) vectors.
By choosing a proper activation function, such as \textit{sigmoid} or \textit{tanh} functions, FCN can achieve outstanding performance on classification problems.
In contrast, CNN accepts higher dimensions like 2D image or 3D vectors.
With the advantages of the convolution operation, it extracts the spatial relationship of the input signal and is expected to achieve learning in long-range interaction by stacking layers sequentially.
Residual Deep Convolutional Neural network (Resnet DCNN) \cite{he2016deep} is well-known to be a more robust alternative to FCNN as the residual function can provide a smooth and stable gradient.
The t-CNN is the inverse operation of the normal convolution process, which is typically introduced in the DNN-based inverse design approach, such as the Generative Adversarial Networks (GAN)\cite{goodfellow2014generative}.

\begin{figure}[ht]
\centering
\includegraphics[width=0.9\linewidth]{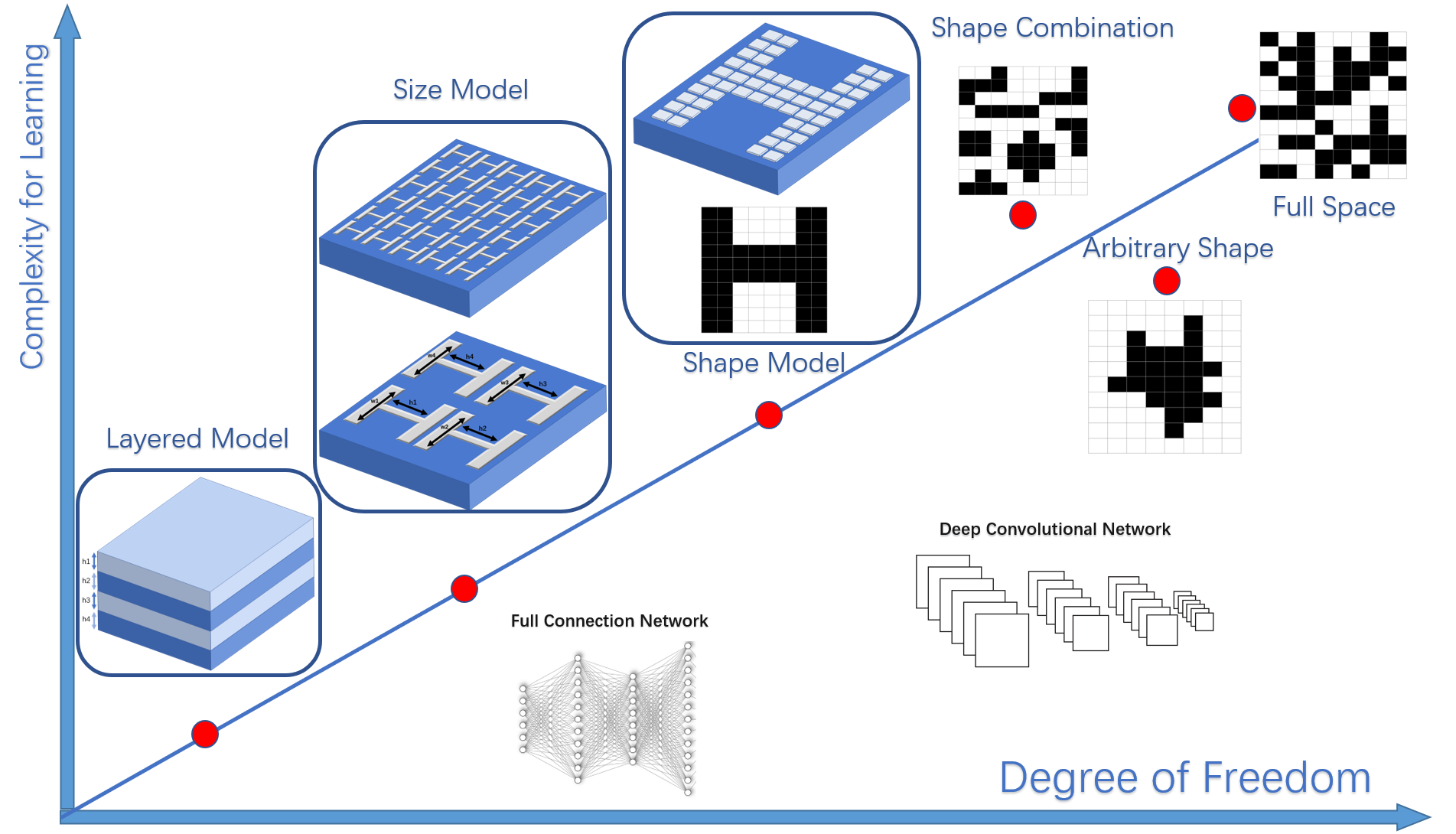}
\caption{Complexity of metasurfaces as a degree of freedom (from left to high: low to high).
The first one (layered model) contains only the thickness per layer.
The second one (size model) includes the surface-size-dependent specific regular patterns.
The third one (shape model) uses a regular shape as a pattern.
The last three are the 3 complex patterns studied in this paper, which are shape combination (PTN), arbitrary connecting shapes (PLG), and full space random pattern (RDN).}
\label{demo_for_metasurface}
\end{figure}

In the following sections, we first provide a short overview of DL based framework used for designing metasurfaces.
The complexity of various types of metasurfaces with different degrees of freedom is shown in Fig.\ref{demo_for_metasurface}.
In this paper, we consider three complex metasurfaces:
(a) Arbitrary connecting polygons (PLG), (b) Basic pattern combination (PTN), and (c) Fully random binary patterns (RDN).
The goal is to explore DL based design for any given complex metasurfaces for a broadband EM response.
In our experiments, the EM response is focused on the reflection of a
broadband EM wave (from 2 to 12 GHz) on the given dataset of complex metasurfaces (PLG, PTN and RDN), where the reflection as a function of frequency can be predicted by using different DL models based on our training procedures for both forward and inverse design.
Successful results and limitations will be evaluated and discussed.
By considering the subordinate relation in these three metasurfaces, we also use a cross-benchmarking to evaluate the ability of the FCNN model in using different metasurfaces in both training and testing.
\hl{Finally, we conclude the paper that MSDCNN is able to provide good performance for each of the three complex metasurfaces studied in this paper, however there is no one single universal DCNN model that is able to perform well for all of them simultaneously. }
Other DL models like the graph neural networks or complex value neural network, will be likely promising candidates for further improvements.

\section{Metasurface design using deep learning}\label{MDCNmodel_and_RIS}
Many successful applications of DL models have been reported in the literature for the design of different metasurfaces \cite{liu2018training,an2020freeform,zhen2021realizing,deng2021neural,cardin2020surface,nadell2019deep, malkiel2018plasmonic,peurifoy2018nanophotonic,an2019deep,tahersima2019deep,asano2018optimization,jiang2019free,jiang2019global,haninverse,sajedian2019finding,liu2018generative,ma2018deep,ma2021deep,jiang2020deep}. \hl{
In metasurface designs, the design parameters and the desired EM responses are of key concerns. The design parameters are properties associated with the metasurfaces such as working frequency and geometrical parameters that describe the physical structure of the metasurfaces. 
These parameters are normally limited in a continuous variable range $R$, so the problem can be abstracted as a mapping between $R^n$ and $R^k$, where $n$ is the number of controllable/designable variables and $k$ is the dimension of desired response/design target.
In this sense, the mapping from $R^n$ to $R^k$ is the forward prediction of metasurface designs, while the mapping from $R^k$ to $R^n$ is the inverse-design.}

Depending on the complexity of the controllable variables, we can categorize them into two types.
The Type-1 design task is of relatively low complexity featuring a template with well prescribed shapes that can be described by a handful of geometrical parameters like thickness, spacing and width.

For $k=1$ case, it is the simplest single-regression task.
In some cases, the design target is a frequency-dependent response ($k >$ 1), and the controllable parameters dimenstion are limited to small $n$.
As an example, a prior work \cite{liu2018training} is focused on the electromagnetic scattering of alternating dielectric thin films with a combination of different thicknesses and materials (also known as the layered model).

For such Type-1 tasks, the fully connected network (FCN) model is sufficient to predict the corresponding EM response which agrees with physics based simulated results (obtained from a numerical solver).
However, it is not suitable for the inverse design task that its training becomes unstable and it is also slow due to the inconsistency of the dataset used.
An encoder-decoder network, known as Tandem \cite{liu2018training} has been introduced to solve this problem, which quickly becomes a popular framework in the metasurface design.
Other improvements includes enhancing the FCN for more robust performance by using deeper networks \cite{malkiel2018plasmonic,peurifoy2018nanophotonic,an2019deep,an2020freeform}.

The Type-2 design task allows a higher degree of complexity, which typically features a 2D metasurface and it can be viewed as a mapping of $R^{n} \times R^{m} \to R^k$.

This task can be converted to the above Type-1 task ($R^{n\times m} \to R^k$) if $n$ and $m$ are small \cite{tahersima2019deep}.
However, when the size ($n$, $m$) is very large, the deep FCN will be too expensive and unstable for computation.
Instead of FCN, convolutional neural network (CNN) provides an distinct advantage for such multi-dimensional inputs.
The convolutional operation in CNN is ideal in capturing the spatial relationship of the inputs.
By using the deeply-stacked convolutional layers, both local and long-range effects are expected to be captured by CNN.
For example, at $k=1$ (prediction of the quality factor of a cavity) \cite{asano2018optimization}, CNN with only four layers is capable of producing outstanding prediction.
For $k=2$, generative adversarial network (GAN) has been successfully applied to design a meta-grating component \cite{jiang2019free}.
Furthermore, GAN has also been incorporated for a physics-driven and data-free neural network \cite{jiang2019global}.
As mentioned above, for Type-1 problem, FCN is found to be effective if the input ($R^m$) and target output ($R^k$) have comparable small dimensions.
However, it is not suitable for the Type-2 problem, the inverse design becomes troublesome due to the complication of dimensionality.
It is reported that GAN can efficiently discover the correct metasurfaces in using user-defined and on-demand spectra as input parameters \cite{liu2018generative}.
A bidirectional neural network is successful in designing three-dimensional chiral metamaterials\cite{ma2018deep}

For a broadband response of metasurface, we may have large $k$ over a large frequency range.
Depending on the resolution requirement and frequency range, the value of $k$ can be 1000 or much more is required.
For this problem, mathematical transformations like Fourier, Wavelet, or simple down-sampling can help to extract the most significant information from the EM response.\cite{RLin20}
\hl{Other methods like the contrast-vector\cite{haninverse} designed to emphasize important features like peak location of the EM response, can also improve the efficiency of the inverse design. } 
With these methods, the dimension from $k=1000$ can be compressed to $k<100$ with good accuracy.

\hl{The above discussion} suggests that different DL models are required depending on the specific types of the metasurfaces and its required EM response.
For complex metasurfaces with irregular patterns (see Fig.\ref{demo_for_metasurface}), we are not sure if DL models will work especially for a broadband EM response, 
\hl{in which} studied domain space is big with large values of $n$, $m$ and $k$.
Thus it is the focus of this paper to evaluate the performance of DL models for the broadband response of such complex metasurfaces.

The remainder of the paper is organized as follows.
In Sec.\ref{MSDCNN}, we will introduce the training procedures to obtain a well-performance \hl{forward prediction} and inverse design model based on the convolution neural network (CNN).
We will report the good prediction of our model for both fast forward and inverse design of PLG based complex metasurfaces.
In Sec.\ref{Cross-banchmark}, we will introduce the other two complex metasurfaces: PTN and RDN.
We will extend the original DL model developed for PLG to PTN and RDN.
Improvements in using deeper DL models and large datasets are reported.
The weak generalization of one unique DL model to all PLG, PTN and RDN is discussed with detailed statistal measure and cross-benchmarking.
Finally, Sec.\ref{Conclusion} concludes with a summary and raises some future prospects in using DL models to deal with the complex metasurfaces.
We argue that complex metasurfaces studied here may serve as a good platform to test the capability or limit of ML in analyzing a complex design problem even though the EM response of the complex metasurface is well governed by the Maxwell equations.

\begin{figure}[ht]
\centering
\includegraphics[width=0.8\linewidth]{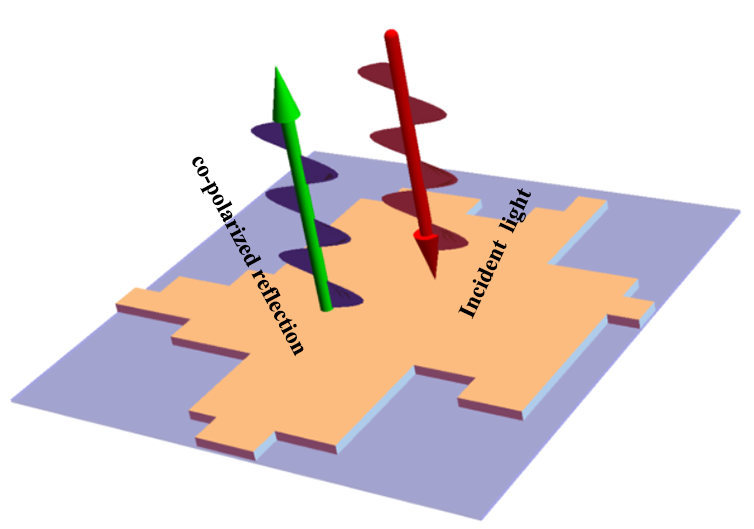}
\caption{The diagram for a co-polarized reflectance (coPR) of a EM wave incident vertically on a purely reflective metasurface.
}
\label{diagram_inject_transmit}
\end{figure}

\begin{figure}[ht]
\centering
\includegraphics[width=0.8\linewidth]{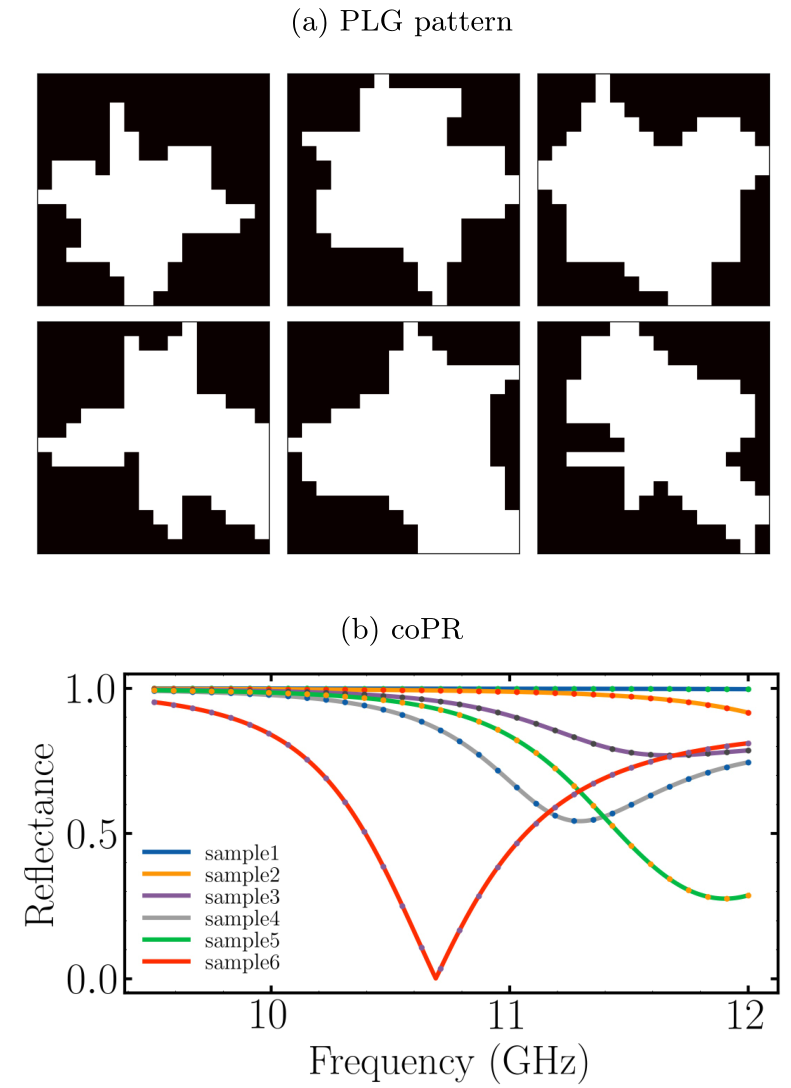}
\caption{PLG dataset:
(a) 6 randomly created arbitrary connecting polygon (PLG) patterns,
(b) The co-polarized reflectance (coPR) calculated from CST for each PLG pattern}
\label{snap_for_PLG_dataset}
\end{figure}

\section{Metasurface design deep convolutional neutral network (MSDCNN)}\label{MSDCNN}

We propose a metasurface design deep convolutional neural network (MSDCNN) framework for both forward-design and inverse-design of complex metasurfaces.
In particular, a co-polarized reflectance (coPR) of a purely reflective metasurface over a frequency range of 2 to 12 GHz is chosen for the purpose of demonstration.
A high quality dataset is important for the training of MSDCNN, thus we rely on automated full wave simulations with F-solver in CST to provide accurate characterization of various complex metasurfaces to the corresponding calculated coPR values.

Each metasurface created in our experiment is represented by a unique pattern encoded by a 16x16 matrix that made up of 0 and 1.
The binary setting of 1 or 0 corresponds respectively to the presence or absence of a square copper patch (0.5 mm x 0.5 mm x 0.018 mm) overlayed on top of a dielectric substrate ($\epsilon_r=2.65 \times (1+0.003i)$ and $\mu_r=1$), which is backed by a 0.18-mm-thick copper plate.
This together with a padding of 1 mm on the sides forms the unit cell used in the CST simulation.
Simulations were performed with unit cell boundary condition in $x$ and $y$ direction and open boundary condition in the $z$ direction.
An $x$-polarized plane wave is incident normally from the top of the metasurface as illustrated in Fig.\ref{diagram_inject_transmit} and the reflection is measured as a function of frequency.
We generate 30,000 samples of arbitrary connecting polygon patterns (PLG) with six examples shown in Fig.\ref{snap_for_PLG_dataset}(a). 
The data are randomly split into a training set of 27,000 samples and a test set of 3000 samples for training purposes. 
The calculated coPR for each PLG has 32 points evenly distributed over a frequency range from 9.5 GHz to 12 GHz as shown in Fig.\ref{snap_for_PLG_dataset}(b). 
\hl{In our experiments, we have found that uniform sampling method is better than fast Fourier transform (FFT) and wavelet transform. 
Having tested several different numbers of sampling points, 32 sampling points has been determined to produce satisfactory outcome with minimal information loss. 
Furthermore, it is to keep the network compact and efficient. 
Increasing sampling points has led to slight increase in network parameters and longer convergence time but without significant improvement in our experiments. }

\begin{figure}[htbp]
\centering
\includegraphics[width=0.8\linewidth]{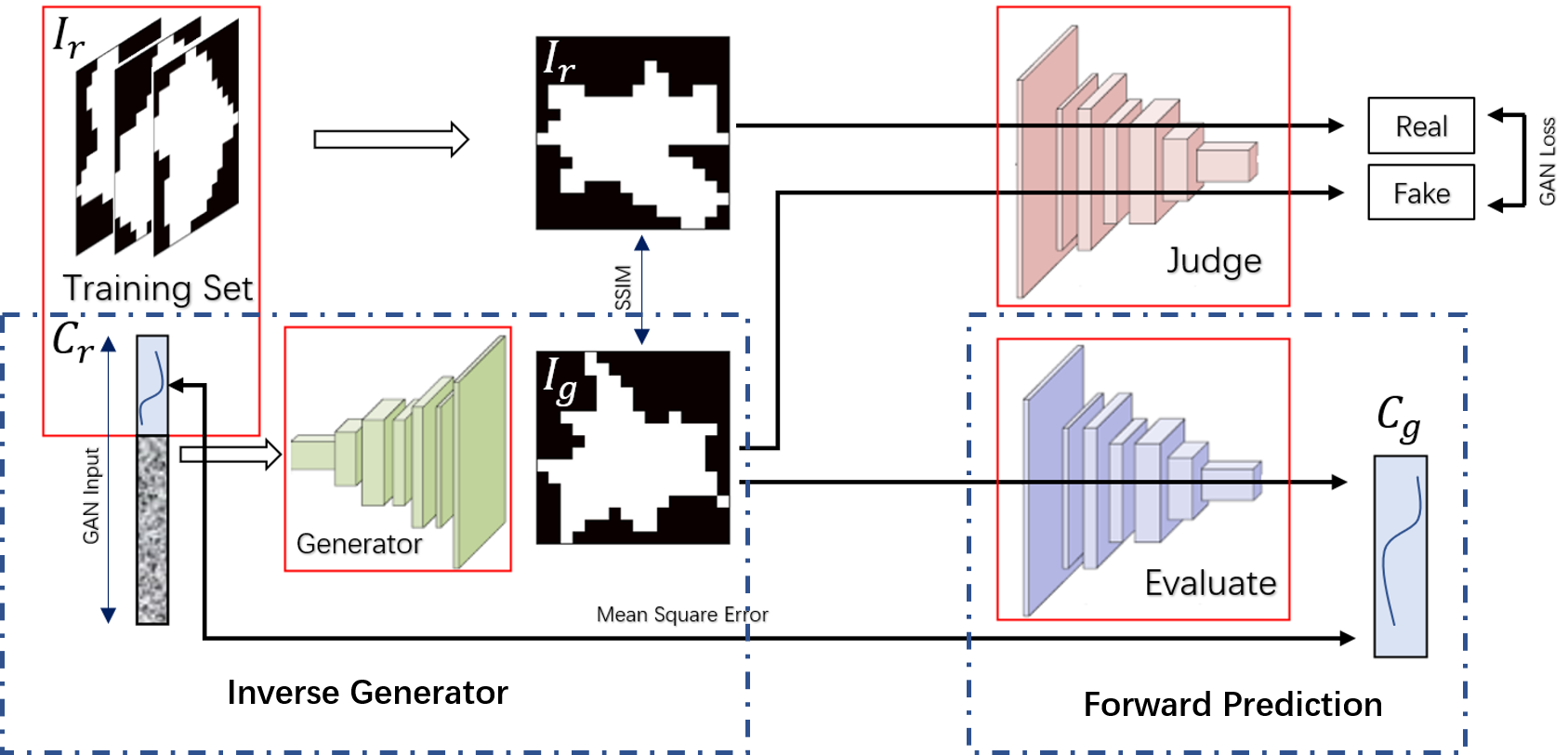}
\caption{Metasurface design deep convolutional neutral network (MSDCNN); \hl{
     $I_r$ is the input pattern (in the form of image) and $C_r$ is the actual EM response associated with $I_r$. 
     $I_r$ and $C_r$ together form the training dataset. 
     The input of the Inverse Generator branch is the actual EM response, $C_r$, appended with a fully random vector. 
     $I_g$ is the generated pattern from the Inverse Generator branch as well as the input to the Forward Prediction branch.
     $C_g$ is the EM response predicted by the Forward Prediction branch which is trained to minimize the mean squared error between $C_g$ and $C_r$.
     }}
\label{whole_workflow}
\end{figure}

\hl{
Our MSDCNN framework in Fig.\ref{whole_workflow} is composed of three branches:
(a) Forward Prediction branch (evaluation of an image to predict a coPR);
(b) Inverse Generator branch (generation of an image from a given coPR); and
(c) Judge branch (measure the agreement of the genereated pattern). }
\hl{Resnet18S is adopted in the evaluation branch which consists of eight Resnet blocks with LeakReLU activation function. It has been adapted from popular Resnet18 model\cite{he2016deep}.}
This evaluation branch predicts the corresponding coPR associated with a input 2D image (like PLG pattern).
The generation branch has five transposed convolution blocks followed by a \textit{tanh} activation function.
This inverse generator branch suggests a 2D image associated with the input EM response like coPR.
The judging branch is made up of a series of regular convolution blocks, which is used to compare the generated image obtained from the generation branch to the input image, in terms of the confidence level of the matching.
A successfully trained \hl{forward prediction branch} is regarded as a  replacement for the numerical solver for the forward design and the \hl{inverse generator branch} is used to perform the inverse design.

In our MSDCNN, we first train a very accurate evaluation model to predict coPR to replace the time-consuming CST simulation.
Secondly, we build a high-quality image generator under the GAN framework in order to convert any random sequences to a 2D polygon-like pattern.
Using the evaluation and generation branches, we can establish a differentiable mapping between the 2D images and the input coPR, so it becomes a useful metric in quantifying the matching between the two inputs.
By enforcing a minimization (or optimization) of this metric in the training procedure, it becomes a pattern-coPR converter.
This training procedure encourages the model to search for the best candidate pattern for the evaluation branch.
Thus the accuracy of the evaluation branch is critical to the performance of the  MSDCNN, which largely depends on the quality and quantity of the dataset used in the training.
More details on the architecture of MSDCNN and the training procedure can be found in Supplementary materials.

\begin{figure}[htbp]
\centering
\includegraphics[width=0.8\linewidth]{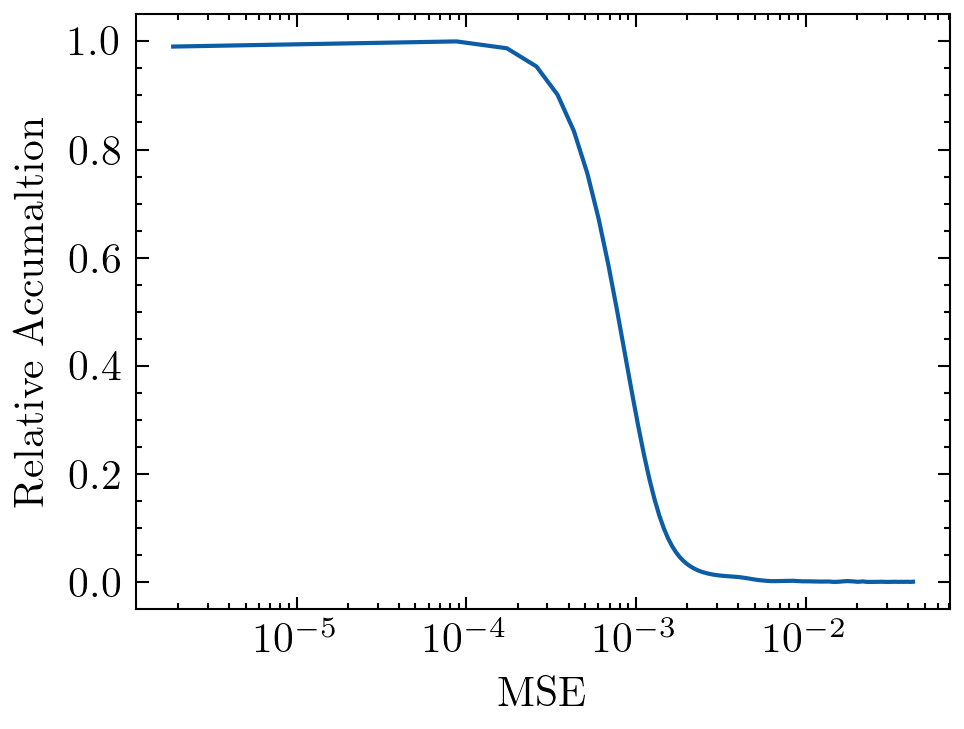}
\caption{Accuracy of the forward design task (normalized by the maximal counting) as a function of MSE
defined as $s=|C_c-C_r|^2$, where the $C_r$ is the true coPR (obtained from CST) and $C_c$ is the predicted coPR by the evaluation branch.
}
\label{PLG_FWD_MSEhist}
\end{figure}

To measure the accuracy of MSDCNN on the PLG patterns, we use the mean square error (MSE).
Our results suggest that the MSDCNN can obtain high accuracy for both the \hl{forward prediction} and the inverse design tasks from the evaluation and generation branch, respectively.
For the forward prediction task, the MSE is defined as $s=|C_c-C_r|^2$, where the $C_r$ is the true coPR (obtained from CST) and $C_c$ is the predicted coPR from evaluation branch in MSDCNN.
Fig.\ref{PLG_FWD_MSEhist} shows the histogram of the accuracy of the 3000 test samples as a function of MSE or $s$.
It is clear that most of MSE are less than $10^{-4}$, which confirms the superior performance of the evaluation branch or forward design of MSDCNN for PLG patterns.

\begin{figure}[htbp]
\centering
\includegraphics[width=0.8\linewidth]{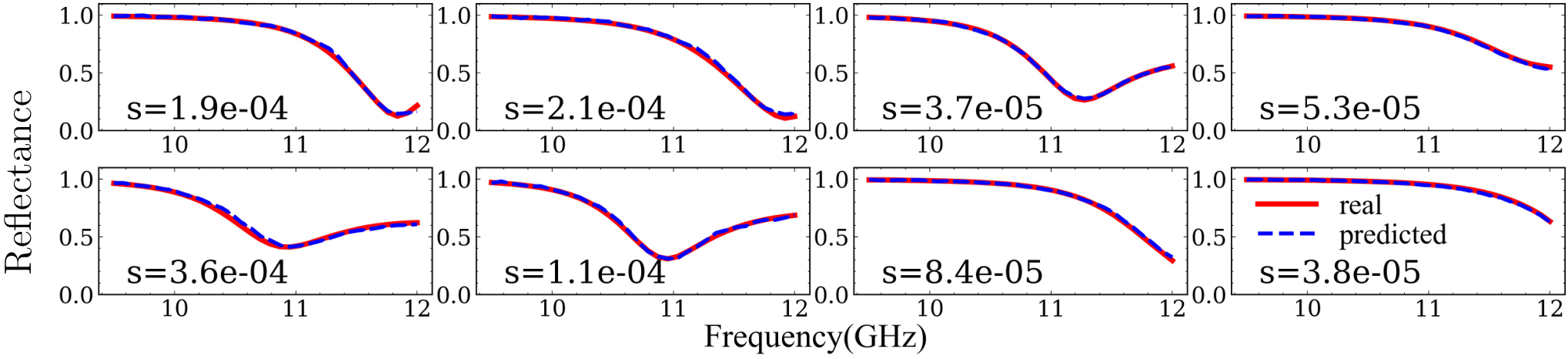}
\caption{The comparison between the real calculated coPR ($C_r$) by CST (solid red lines) and the predicted coPR
($C_c$) by the evaluation branch of 8 random PLG patterns. The MSE between them are indicated as $s=|C_c-C_r|^2$.}
\label{demo_for_PLG_FWD_result}
\end{figure}

\hl{In Fig.\ref{demo_for_PLG_FWD_result}, eight randomly selected test cases
are plotted to demonstrate the excellent agreement between predicted and true calculated coPR.}
More importantly, the model can correctly capture both the locations and magnitudes of the peaks or variation in the spectrum from 9.5 to 12 GHz.
Here we have coPR = 1 at lower frequency from 2 to 9.5 GHz (not shown).
The average error of the evaluation branch is $s= 3.7 \times 10^{-4}$, thus confirming that the evaluation branch has been successfully trained with high accuracy and thus can be used as a fast computational tool to replace the traditional EM simulator for the design of complex PLG metasurfaces.

\begin{figure}[ht]
\centering
\includegraphics[width=0.8\linewidth]{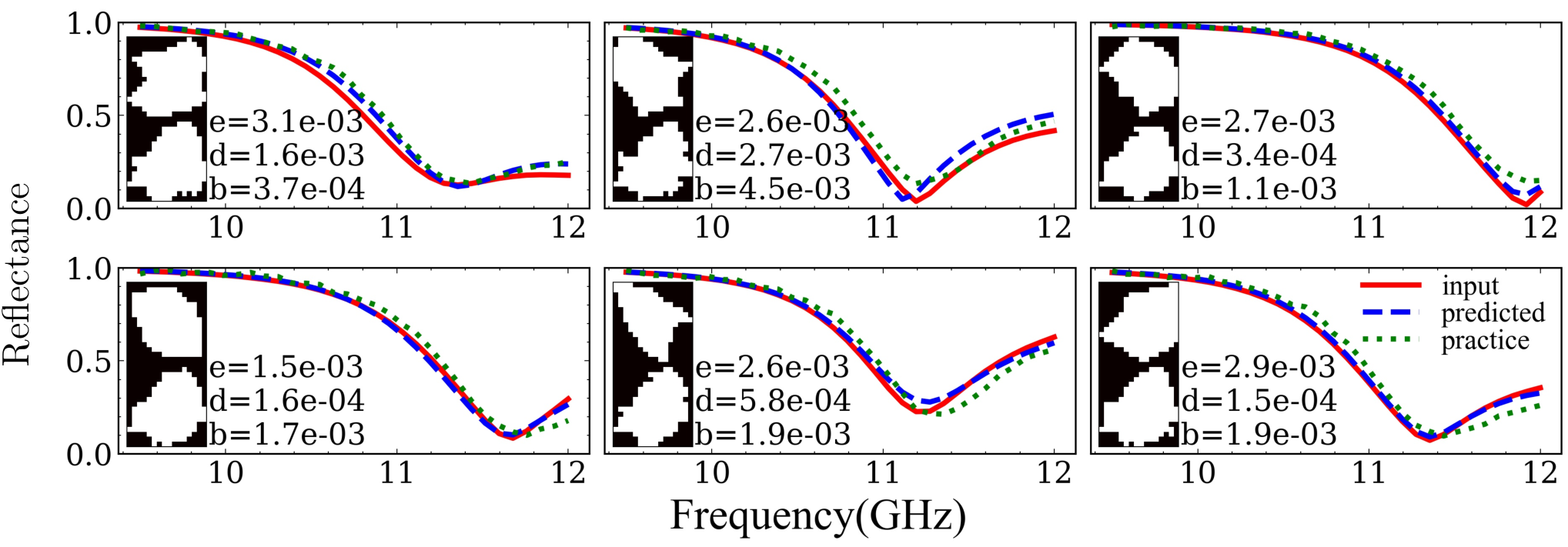}
\caption{\hl{Inverse design for six PLG patterns. The incident electric field is x-polarized. The x-axis for each image is frequency from 2GHz to 12GHz, the y-axis is the reflectance ranging from 0 to 1. Every figure contains the input or the desired coPR $C_r$ (red), the
estimated coPR of the predicted image (by inverse design) $C_g$ (blue), and the CST calculated coPR $C_p$ (green) of the predicted image.
The MSE are $e=|C_r-C_p|^2$, $d=|C_r-C_g|^2$, and $b=|C_g-C_p|^2$.
Upper (lower) image is, respectively, the input image and the predicted image by the generation branch.}}
\label{demo_for_GAN_result_on_PLG}
\end{figure}

For inverse design, the generation branch suggests a corresponding 2D image of a PLG-like metasurface for a given coPR spectrum.
There are 3 metrics used to determine the accuracy:
$C_r$ is the input (or real) coPR
$C_g$ is the predicted coPR from the evaluation branch, and
$C_p$ is the actual coPR computed by CST based on the 2D image created from generation branch.
An ideal good inverse design demands a low error between $C_r$ and $C_p$ determined by $e=|C_r-C_p|^2$.
However, $C_p$ is calculated from on the time-consuming full wave simulation and avoiding such lengthy simulations will largely reduce the training time required in our model.
Thus, we choose to optimize the the alternative error between $C_r$ and $C_g$, which is defined as $d=|C_r-C_g|^2$.
Finally, the error between $C_p$ and $C_g$ is $b=|C_g-C_p|^2$.
In Fig.\ref{demo_for_GAN_result_on_PLG}, we show the comparison of 6 cases with their respective values of $e$, $d$, and $b$.
The results show that the $C_r$, $C_p$ and $C_g$ agree well with only small errors in the range of $10^{-3}$ to $10^{-4}$.
This indicates that the generation branch has been successfully trained with the ability to provide promising inverse design satisfying the input coPR spectrum.
\hl{
It is observed that the designs produced by the generator are not the same with the reference upper image. 
This is due to the patterns and the EM responses not having one to one mapping. 
In this case, the network will provide a design with the closest possible EM response in the context of training dataset. 
Furthermore, we have used a well-trained evaluate branch as a part of GAN rather than directly employing a full wave simulator to compute mapping from EM response to pattern. 
Those technologies will help expand the expression capability of the model and overcome the data inconsistency\cite{peurifoy2018nanophotonic}.}

The above findings shown in Fig.\ref{demo_for_PLG_FWD_result} and Fig.\ref{demo_for_GAN_result_on_PLG} have proved the possibility of using MSDCNN in complex metasurface design such as PLG patterns.
Firstly, the evaluation branch (forward design) can provide an accurate prediction of a broadband EM response from 2 to 12 GHz.
Secondly, the generation branch can suggest corresponding PLG based metasurfaces to satisfy the input broadband EM response.
In the following section, we will extended this capability to other types of complex metasurfaces, in order to assess the broader performance of MSDCNN and to understand its limitation.

\section{Extension to Other metasurfaces}\label{Cross-banchmark}
For well-known image datasets in computer science like MNIST (hand written digital database), CIFAR and ImageNet, the DL algorithm can achieve state-of-the-art performance on many tasks like recognition, segmentation, and tracking.
Those datasets generally contain prior knowledge or characteristics based on human cognition, and the learning target is the patterns attributed with color, shape, and position.
Such a pattern is typically polygon-like or, more precisely, a connected manifold.
Unlike these images for human recognition, the complex metasurfaces learning are governed by physics like Maxwell equations.
Few studies has been devoted to the effectiveness in using DL to predict the underlying physics-based outputs concealed in such complex metasurfaces.
Our initial success on the aforementioned PLG dataset (in previous section) intrigues us to expand the capability to accommodate other complex patterns with similar high degree of freedom to study the effectiveness of the proposed MSDCNN framework.

\subsection{Three datasets: PLG, PTN and RDN}\label{three_datasets}
\begin{figure}[ht]
\centering
\includegraphics[width=0.8\linewidth]{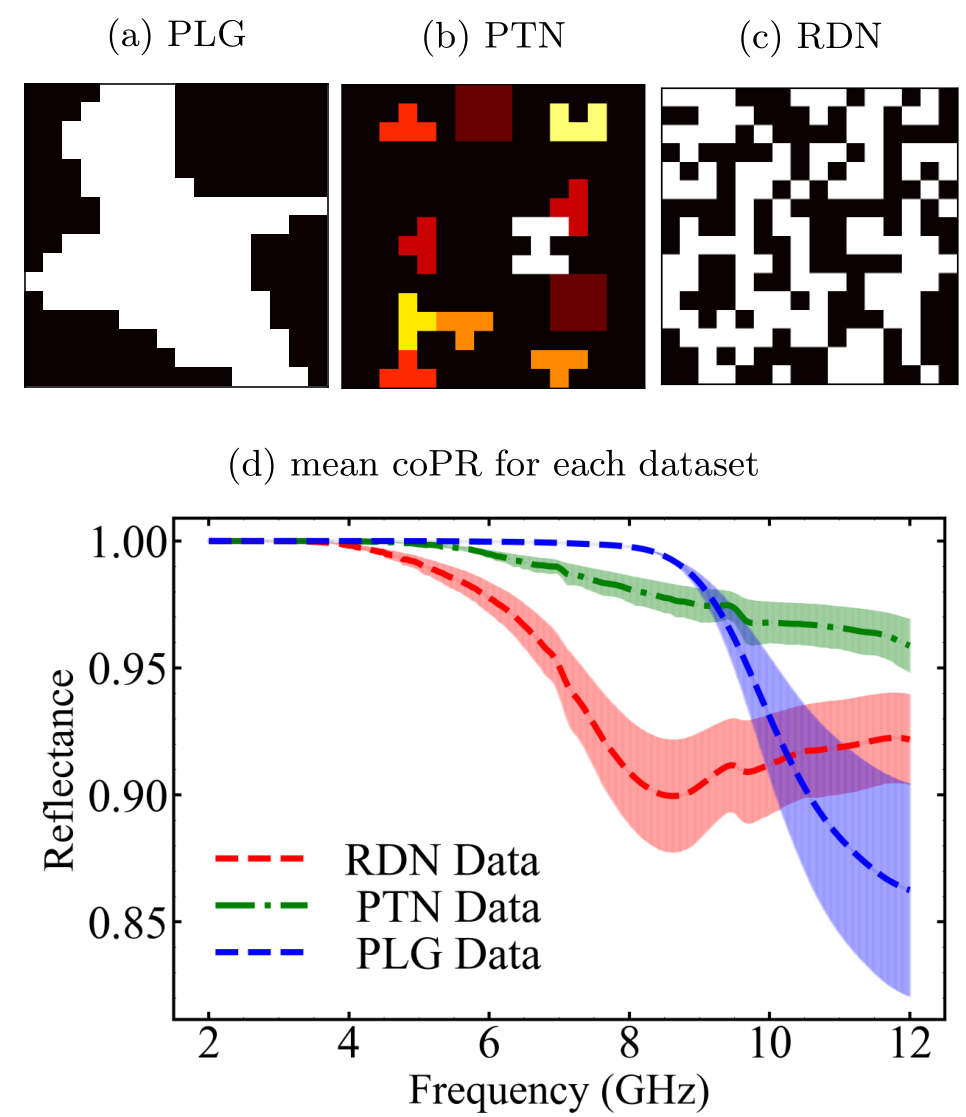}
\caption{Images of PLG, PTN and RDN with their respective mean and variance of coPR calculated by CST as a
function of frequency. Note the PTN image (in b) is colored for different basic shapes (for illustration only).}
\label{dataset_and_its_property}
\end{figure}

The two other two types are pattern-combination (PTN) and the random (RDN) datasets.
Including PLG, we have 3 types of complex metasurfaces [see Fig.\ref{dataset_and_its_property}]:
(a) Arbitrary connecting polygons (PLG), (b) Basic pattern combination (PTN), and (c) Fully random binary patterns (RDN).
Note the patterns in all 3 metasurfaces are encoded into a binary matrix of size $16\times16$ ($Z_2^{16\times16}$).
The PLG pattern requires connectivity, which are common in manufacturing  design \cite{haninverse,liu2018generative,sajedian2019finding,an2020freeform}.
The PTN pattern is the combination of some basic shapes like square (9 pixels), cross (5 pixels), triangle (4 pixels) with four directions, U-shape (5 pixels), and H-shape (7 pixels).
The RDN pattern is the fully random pattern with no constraint.
The coPR of each created pattern is calculated by CST as a function of frequency from 2 to 12 GHz
The statistics of the calculated coPR (mean and variance) for each dataset is shown in Fig.\ref{dataset_and_its_property}(d).
At low frequency, coPR is 1 (perfect reflection).
For PLG shape, we have perfect reflection at frequency lower than 9.5 GHz, thus only the limited range from 9.5 to 12 GHz are shown in previous Figs. \ref{demo_for_PLG_FWD_result} and \ref{demo_for_GAN_result_on_PLG}.

Among the three datasets, RDN has the largest domain or highest degree of freedom in complexity.
Thus it is interesting to know if a well-trained DL model based on the RDN dataset could be able to function well on  different patterns like PTN and PLG datasets.
Thus it is desirable to extend the MSDCNN framework (previously trained on PLG dataset) to PTN and RDN dataset.
Good performance can be obtained for all 3 datasets with modification of the DL models used or with larger sample size (see below).
The generalization of only one unique DL model for all datasets is challenging.

Similar to PLG dataset, the size of training and test sets for RDN and PTN dataset is 27,000 and 300 respectively, unless it is mentioned otherwise.
In some experiments, the number of RDN dataset is increased to 108,000 for better performance.
Other than the standard Resnet18S model used, we also apply other DL models such as Resnet34 and ResNa.
In our comparison, the performance of generation branch heavily depends on the accuracy of the evaluation branch.
Therefore, the discussion below will be centered around the performance of the evaluation branch (forward design) to predict the coPR for a given arbitrary metasurface.
The performance is evaluated not only by the MSE but also other statistical measurements, such as mean, variance, and kurtosis.

\subsection{Improvement and Limitation}
DCNN is well-known for its great generalization in computer vision tasks.
For example, Resnet18 can perform well for most computer vision tasks.
Since our MSDCNN model is a variant of DCNN and with the high similarity between our complex metasurfaces and digital images (used in computer vision), we speculate that the success of MSDCNN on the PLG dataset may work on the PTN dataset and even the RDN dataset.

\begin{figure}[htbp]
\centering
\includegraphics[width=0.8\linewidth]{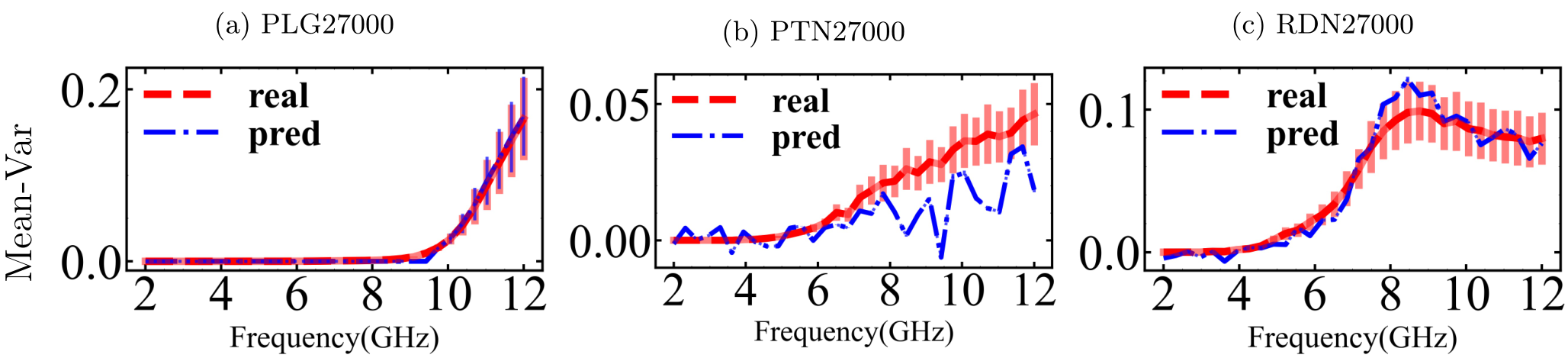}
\caption{The mean-variance of the real CST calculated values of $1 -$ coPR (red) and the prediction (blue) by same
MSDCNN-Resnet18S for all datasets:
(a) PLG,
(b) PTN and (c) RDN.
Note (a) PLG case has outstanding matching.
The (b) PTN (b) and (c) RDN cases show poor agreement.
}
\label{Error_Ku_Bar_for_27000_1}
\end{figure}

\begin{figure}[htbp]
\centering
\includegraphics[width=0.8\linewidth]{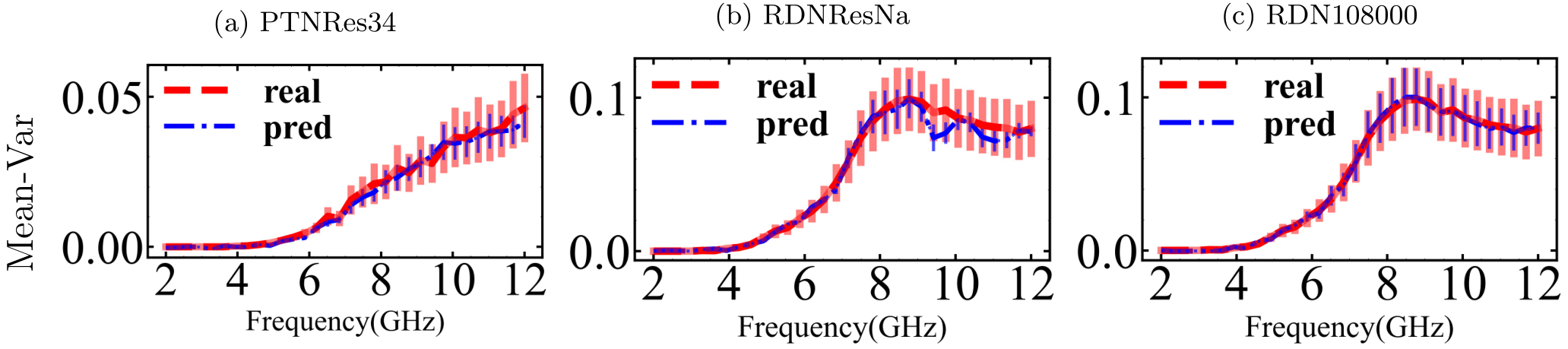}
\caption{The mean-variance of the real CST calculated values of $1 -$ coPR (red) and the prediction (blue) by using
different DL models and datasets:
(a) The Resnet34S model trained on the PTN dataset (27k samples) shows better agreement than Fig.\ref{Error_Ku_Bar_for_27000_1}(b).
(b) The ResNa model trained on the RDN dataset (27k samples) shows better agreement than Fig.\ref{Error_Ku_Bar_for_27000_1}(c).
(c) The MSDCNN-Resnet18S model trained on RDN dataset with more samples (108k) has better agreement than both Fig.\ref{Error_Ku_Bar_for_27000_1}(c) and Fig.\ref{Error_Ku_Bar_for_27000_2}(b).
}
\label{Error_Ku_Bar_for_27000_2}
\end{figure}

\begin{figure}[htbp]
\centering
\includegraphics[width=0.8\linewidth]{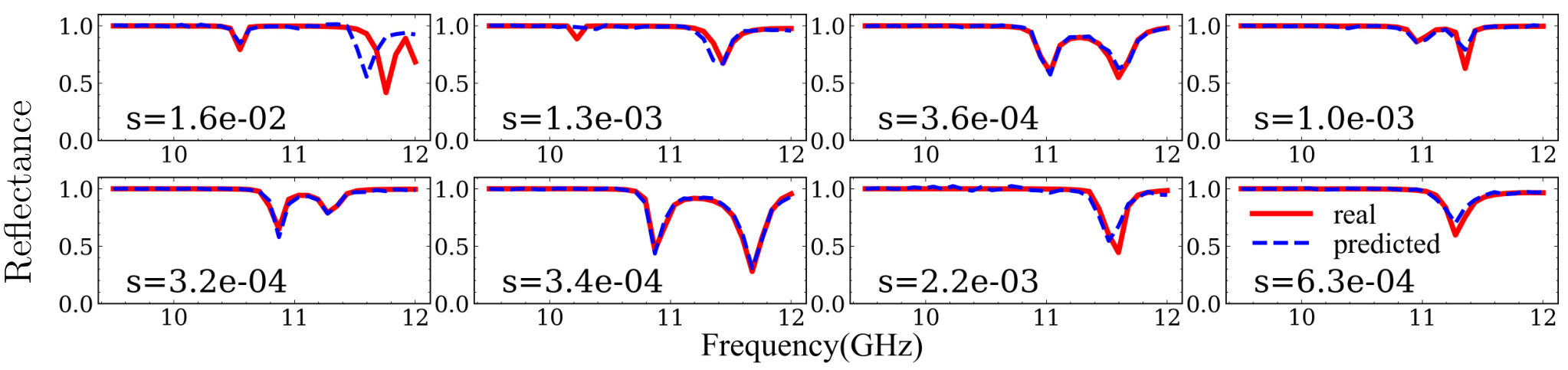}
\caption{Comparison of 8 randomly selected cases from RDN108000 datasets with different mean square error: $s=|C_c-C_r|^2$, where $C_r$ is the real CST calculated coPR and $C_c$ is the DL predicted coPR.}\label{RBDemo_result_for_the_RDN108000_dataset}
\end{figure}

\begin{figure}[htbp]
\centering
\includegraphics[width=0.8\linewidth]{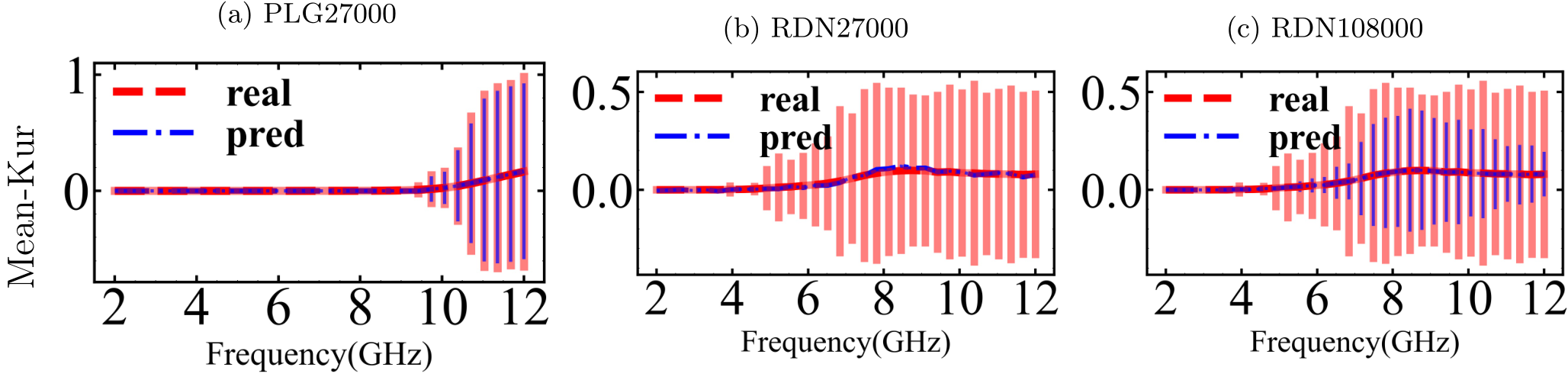}
\caption{The mean and kurtosis values of MCDCNN-Resnet18S for 3 different trained sets:
(a) The well-trained PTN dataset has perfect kurtosis matching.
(b) The not well-trained RDN dataset (27k samples) falls into the naive local minimum.
(c) The better-trained larger RDN dataset (108k samples) avoids partially to fall into naive local minimum.
}
\label{Ku_Bar_for_27000_108000}
\end{figure}

\begin{figure}[htbp]
\centering
\includegraphics[width=0.8\linewidth]{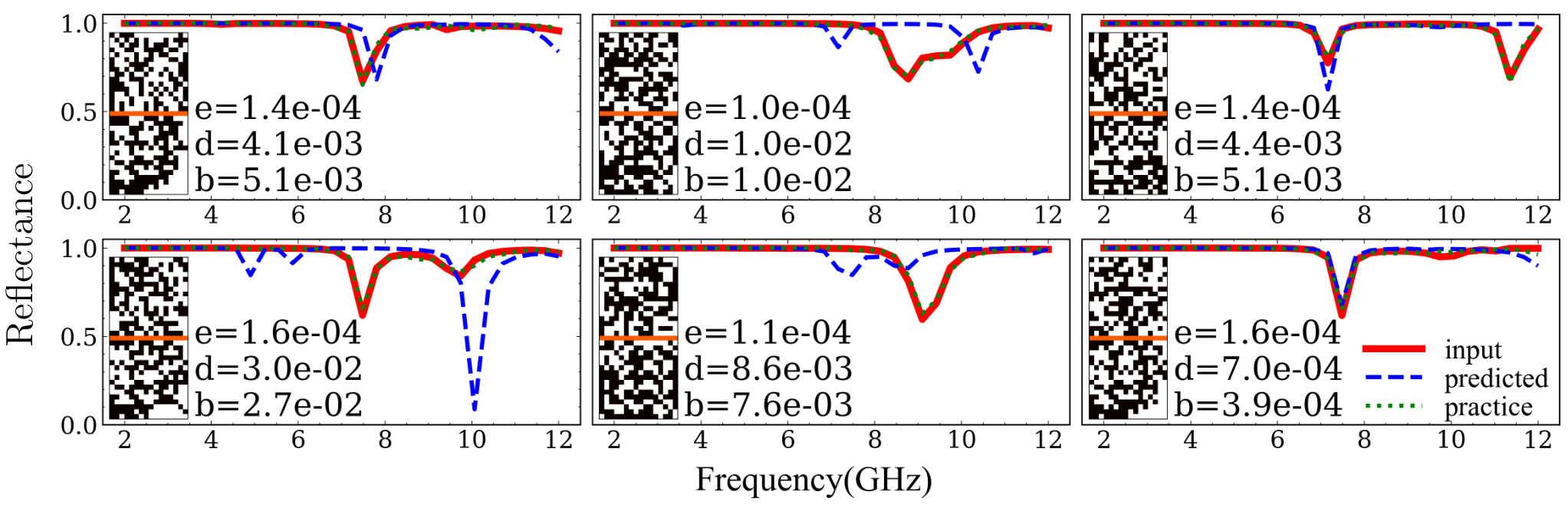}
\caption{Inverse design for 6 random examples from RDN108000 model. Every figure contains the input coPR $C_r$ (red), the estimated coPR of the predicted image (by inverse design) $C_g$ (blue), and the CST calculated coPR $C_p$ (green) of the predicted image.
The MSE are $e=|C_r-C_p|^2$, $d=|C_r-C_g|^2$, and $b=|C_g-C_p|^2$.
The upper image is the input image and the lower one is the predicted image by the generator branch (inverse design).}
\label{GAN_result_RDN108000CURVE}
\end{figure}

While finding one unique and universal DL model for all $arbitrary$ complex metasurfaces is hard, we will show very good improvements can be obtained if we only focus separately on one type of complex metasurface.
This implies that DL model is a useful approach for the design of complex metasurfaces if a particular type can be specified separately to create a suitable DL model.
Fig.\ref{Error_Ku_Bar_for_27000_1} shows the predicted values (blue) by using Resnet18S trained on PLG, PTN and RDN datasets (27,000 samples each), which are labelled as PLG27000, PTN27000 and RDN27000.
The real calculated values from CST are also plotted (red) for comparison.
Notice that the PTN results show the poorer agreement than PLG even same number of samples are used, which suggests that the Resnet18S (good for PLG) is not necessarily good for the PTN dataset.

Deeper/better DL models or larger dataset can probably enhance performance.
Fig.\ref{Error_Ku_Bar_for_27000_2} demonstrates these efforts with significant improvements.
The details of the adapted deeper ML models can be found in the supplementary materials.
The Resnet34S model is the deeper version of the Resnet18S which has more parameters and greater capacity.
As expected, the performance of Resnet34S is better in training the PTN27000 dataset as compared to Resnet18S, where the agreement in Fig.\ref{Error_Ku_Bar_for_27000_2}(a) of PTNRes34 is better than Fig.\ref{Error_Ku_Bar_for_27000_1}(b).
ResNa model combines CNN with long short-term memory (LSTM) network and it has less layers and fewer parameters than Resnet18S.
ResNa is expected to handle a more complex pattern like RDN dataset.
Thus the results of RDN27000 in Fig.\ref{Error_Ku_Bar_for_27000_1}(c) has been improved by using ResNa model as shown in Fig.\ref{Error_Ku_Bar_for_27000_2}(b).
By expanding the sample size of RDN dataset (based on Resnet18S) from 27000 to 108000 (RDN27000 to RDN108000), we also improve the performance as shown in Fig.\ref{Error_Ku_Bar_for_27000_2}(c) in comparison to Fig.\ref{Error_Ku_Bar_for_27000_1}(c).
The improvements reported in Fig.\ref{Error_Ku_Bar_for_27000_2} confirms that good performance can be achieved by using better DL models and/or large training sets for all 3 complex metasurfaces over a frequency range from 2 to 12 GHz.
For completeness, we show 8 random patterns selected from the RDN108000 dataset in Fig.\ref{RBDemo_result_for_the_RDN108000_dataset} to illustrate their small $s$ (mean square error).
Note this good performance is due to same type of complex metasurface used in both training and testing.

Upon further analysis using the high-order statistical measure like kurtosis, we notice that the previous analysis based on mean and variance does not reflect the matching fully.
Note kurtosis is a statistical measure to define how heavily the tails of a distribution differ from the tails of a normal distribution, which helps to determine whether the tails of the distribution containing extreme values.
The calculated mean and kurtosis are shown in Fig.\ref{Ku_Bar_for_27000_108000} for 3 cases: (a) the well-trained PLG27000, (b) the under-trained RDN27000, and (c) the large training set of RDN108000.
The well trained PLG27000 model not only match the variance well in Fig.\ref{Error_Ku_Bar_for_27000_1}(a) but also has a matching kurtosis as shown in Fig.\ref{Ku_Bar_for_27000_108000}(a).
The RDN108000 case although have a good improvement in variance in Fig.\ref{Error_Ku_Bar_for_27000_2}(c) in using larger sample sizes, the disagreement in kurtosis remains significant as shown in Fig.\ref{Ku_Bar_for_27000_108000}(c).
Such disagreement in kurtosis will lead to high bias error in inverse design.
For example, Fig.\ref{GAN_result_RDN108000CURVE} shows the performance of generation branch (or inverse design) when the model is trained on the RDN dataset of 108k samples.
The CST calculated coPR $C_p$ (green) of the predicted image by the generation branch agree well with the desired input coPR $C_r$ (red), but not necessarily in good agreement with the estimated coPR $C_g$ (blue) from the evaluation branch using predicted image.
This finding reveals that MSDCNN-Resnet18S model has high bias error in the generation branch (inverse-design) despite having superior performance in the evaluation branch (forward design).

\subsection{Cross benchmarking between the models}
Consider two datasets A and B, where the domain of B is a subset of the domain of A.
A central question in this section is whether a well-trained model on A will perform considerably good on B.
For example, we are interested to know if the model trained under RDN dataset is good for testing PLG or PTN datasets.
Another question is to verify if CNN based model will perform better than other models such as random forest regressor (RFR), where RFR is an ensemble learning method by constructing many decision trees at the training, and the output is based on the average prediction of all trees.
Note RFR is a general purpose yet powerful model that requires less resources to train and less parameters to be tuned in comparison to CNN based models.
This section focuses on answering these questions by performing cross benchmarking using different training datasets and models.

\begin{figure}[htbp]
\centering
\includegraphics[width=0.8\linewidth]{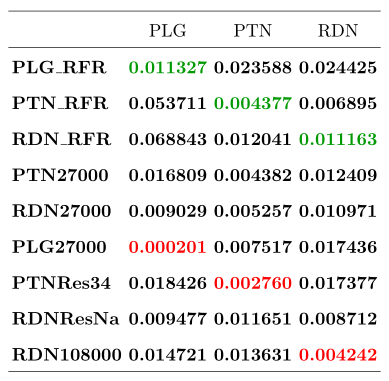}
\caption{(a)(b)The poor performance of the trained RDN108000 on PLG and PTN.(c)The RDN model trained as a function of RDN samples from 80k to 108k and their testing performance on RDN (left axis, about 4X smaller in magnitude), PTN and PLG (both on right axis).}
\label{ErrorBarvar_Test_Result_RDN108000_PLG_PTN}
\end{figure}

Table.\ref{Result_Cross_test} shows the MSE performance of different models, which are trained independently on PLG, PTN, and RDN dataset and evaluated on all 3 datasets (transfer learning).
For example, the row of PLG27000 shows the results using PLG dataset (27k samples) to train MSDCNN-Resnet18S, but the model is tested on the all 3 datasets (PLG, PTN and RDN).
The results marked in red are the best performance (smallest MSE), and the best performance remains on using the same type of dataset for both training and testing.
This finding suggests that the models trained on dataset with larger domain do not necessarily perform better on other datasets even with smaller domain of less complexity
For example, the models trained on higher-level dataset (RDN) do not perform well at lower-level dataset (like PTN and PLG).
By comparing between the RDN108000 (4X more samples) and RDNResNa/PTN27000, it tends to show weaker performance than an under-fit model.
This situation can be regarded as another type of over-fitting on the large domain level, suggesting that the DL-based model is more like a curve-fitting process than learning the physics behind it, which will otherwise allow equivalent performance in using large datasets in training (see move in discussion below).
The weak agreement of the model trained on RDN108000, which are tested on PLG and PTN can be found in Fig. \ref{ErrorBarvar_Test_Result_RDN108000_PLG_PTN}(a) and \ref{ErrorBarvar_Test_Result_RDN108000_PLG_PTN}(b).

To completely investigate the compatibility on the low-level dataset, we need to check the dependence of sample size in the RDN dataset used in the training, which is plotted in Fig.\ref{ErrorBarvar_Test_Result_RDN108000_PLG_PTN}(c) from 80000 to 108000 samples.
The figure shows that, although the performance on the RDN dataset increases (smaller MSE) via more samples, its performance in testing on PLG and PTN's scores are not affected by the trained sample size.
It confirms that the a well-trained model on RDN dataset will not work for other datasets like PLG and PTN even it has small domain.
Finally, in the table, we can conclude that the best well-trained CNN models (red) will have better performance than the well-trained RFR (green), which demonstrates the advantages of using CNN based algorithms.

\begin{table}[ht]
\centering
\begin{tabular}{lccc}
\hline
\textbf{}          & PLG                                      & PTN                                      & RDN                                      \\ \hline
\textbf{PLG\_RFR}  & {\color[HTML]{009901} \textbf{0.011327}} & \textbf{0.023588}                        & \textbf{0.024425}                        \\
\textbf{PTN\_RFR}  & \textbf{0.053711}                        & {\color[HTML]{009901} \textbf{0.004377}} & \textbf{0.006895}                        \\
\textbf{RDN\_RFR}  & \textbf{0.068843}                        & \textbf{0.012041}                        & {\color[HTML]{009901} \textbf{0.011163}} \\
\textbf{PTN27000}  & {\textbf{0.016809}}                      & {\textbf{0.004382}}                      & {\textbf{0.012409}}                      \\
\textbf{RDN27000}  & {\textbf{0.009029}}                      & {\textbf{0.005257}}                      & {\textbf{0.010971}}                      \\
\textbf{PLG27000}  & {\color[HTML]{FE0000}\textbf{0.000201}}  & \textbf{0.007517}                        & \textbf{0.017436}                        \\
\textbf{PTNRes34}  & \textbf{0.018426}                        & {\color[HTML]{FE0000} \textbf{0.002760}} & \textbf{0.017377}                        \\
\textbf{RDNResNa}  & \textbf{0.009477}                        & \textbf{0.011651}                        & \textbf{0.008712}                        \\
\textbf{RDN108000} & \textbf{0.014721}                        & \textbf{0.013631}                        & {\color[HTML]{FE0000} \textbf{0.004242}} \\ \hline
\end{tabular}
\caption{The left column is the model's name.
The first three characters show the training dataset (PTN, PLG, or PTN).
The last characters show the model type, if the type is number means it is the default Resnet18S.
The second, third, and fourth columns show the MSE score on the PLG, PTN, and RDN testing dataset accordingly.
The valid baseline score for RFR is colored by green.
The best model score for DL algorithms is colored by red.}\label{Result_Cross_test}
\end{table}
\printtables

\subsection{Discussion}
Our experiments conducted above have shown that the MSDCNN framework can work well separately for 3 complex metasurface patterns: PLG, PTN and RDN.
It can be a useful and fast computation tool to design complex metasurfaces if different DL model is used specifically for different patterns. 
\hl{Having tested a variety of popular DCNN architectures,  there is NO one unique and universal DCNN model that is suitable for all 3 types of metasurfaces.}
The comparison between the PTN and PLG dataset shows that different complex metasusrfaces require different DL models even using the same number of samples in training.
The comparison between different sample sizes (from RDN27000 to RDN108000) indicates that larger sample size, as expected, is an essential but not the sole factor for the performance.
The poor performance in using a more general dataset with large domain (like RDN) to test on more constrained patterns with smaller domain (like PTN and PLG) suggests that the patterns emerged within a dataset and recognized by DL models are specifically relevant to the individual dataset.
However, such patterns are not generalized enough to constitute an understanding to the governing laws due to physics, namely Maxwell equations.
This phenomenon is similar to the mode collapse issue encountered in GAN model.
An explanation is due to the limitation of using mean-square error (MSE) over the the entire training dataset, and local-minimum-solutions are not captured properly.
The local minimum problem can be avoided by introducing a regularization term in most cases.
Having tried this approach multiple times, we observed that this issue happens frequently when the size of dataset is small despite careful tuning of hyper parameters, which can be attributed to the unsuitable architecture of the DCNN models used in the evaluation branch.

The notion of inductive bias has been proposed \cite{battaglia2018relational} to emphasize the relation between task symmetry and operation ability.
\hl{From this perspective, the datasets like PLG and PTN have locality and spatial symmetry, which is suitable for convolutional operation.
The PLG pattern is also denser with more continuity than the PTN pattern, so it is expected to perform better.
Thus, our MSDCNN-Resnet18S works well on PLG despite a significantly smaller dataset is used, as shown in
Figs. \ref{demo_for_PLG_FWD_result}, \ref{demo_for_GAN_result_on_PLG}, and \ref{Error_Ku_Bar_for_27000_1}(a).}
By improving the capacity of Resnet18s to Resnet34S, it offers better performance on PTN dataset with the same number of samples [see Fig.\ref{Error_Ku_Bar_for_27000_2}(a)].
The most general dataset (RDN) with the highest degree of freedom turns out to be the most challenging model to be trained well. We have to increase the number of samples from 27k to 108k (4 times more) to improve the accuracy to a comparable level [see Fig.\ref{Error_Ku_Bar_for_27000_2}(c)].
From the cross-benchmarking results, the model trained in the larger domain of datasets (like RDN) is not applicable for a smaller or more specific domains like PLG and PTN.
This suggests that a model trained on two datasets of distinct properties are not compatible, even if the domains of the two datasets are of subset relationship.
Thus there may exist a compatible state of the specific ML model that we have not achieved yet, which will require future studies.
The ultimate goal to a unique and universal DL model for the forward and inverse design of arbitrary complex metasurfaces will require new DL models in future investigations.

\section{Conclusion}\label{Conclusion}
In summary, we propose metasurface design deep convolutional neural network (MSDCNN) to study the performance of CNN based models in order to perform the design (both forward and inverse) of complex metasurfaces for broadband electromagnetic (EM) wave reflection from 2 to 12 GHz.
Having experimented on three different complex metasurfaces with high degree of freedom: arbitrary connecting shapes (PLG), basic pattern-combination (PTN) and fully random binary shapes (RDN), it is confirmed that MSDCNN can provide a promising tool (faster than traditional numerical EM solver, such as \emph{CST Studio Suite} EM analysis software) for such complex metasurfaces.
Among them, the best performance is on PLG like metasurfaces, which requires the least efforts to achieve good performance.
In contrast, more advanced ML algorithm is required for PTN while substantially larger sample size is required for RDN in order to achieve the same performance.
\hl{There is no one unique and universal DCNN model that can work well for all of them, thus transfer learning of DCNN between them is not good.}
Such behavior is likely caused by the fact that the sequential stacking of convolution operation is suitable for parsing information which contains spatial locality and highly nearby correlation pattern like natural images.
However, for arbitrary complex metasurface, the information extracted by pure CNN are not sufficient and can be further enhanced.
This is revealed by the performance improvement after combining RNN, demonstrated by ResNa model.
The finding suggests that new DL models are required to achieve a universal DL model for arbitrary complex metasurfaces.
The inductive bias of arbitrary meta-surface design contains not only the simple spatial invariance but also the complex physics governed by the Maxwell equation like the long-range and time-space interaction and non-local interaction, that the current DL models fail to capture according to our studies.
For future works, other models such as graph neural networks, complex value neural network and physics inspired CNN \cite{PINN2019} could be explored as possible candidates for further improvements.

\hl{It is important to note that the focus of this paper is due to the curiosity in understanding the limits of MSDCNN in predicting the EM response of complex and random metasurfaces without specific applications in mind. 
We also ignore any experimental constraints and feasibility of such complex metasurfaces in any applications, which will require future explorations.
However, some complex metasurfaces of such kind  \cite{zhang2017shaping,Jin2018,Guan2020,Wu2020,Chen2016,Li2017,Moccia2017,Yan2015,Gao2015,Liang2015} have been realized and reported in the literature with feature size down to nanometers.  
These metasurfaces are designed using binary coding with some of them \cite{zhang2017shaping,Liang2015,Gao2015,Yan2015} appear to be atypical, unstructured, and random. 
The simple EM reflection problem is for demonstration purpose, the results can certainly be extended to include other effects like incident angles, polarization and other, which can be included using CST for data collection.
For realistic applications, one may need to conduct sufficient experimental data for benchmarking and if possible as training dataset as well.
}

\section{ACKNOWLEDGEMENT}
This work is supported by USA Office of Naval Research Global (N62909-19-1-2047).
T.Z acknowledge the support of Singapore Ministry of Education PhD Research Scholarship.
Y.S.A. acknowledge the support of SUTD Start-Up Research Grant (SRT3CI21163).

\section{SUPPLEMENTARY MATERIAL}
See Supplementary material for the model structure of the deep learning used in the paper.

\section{Data Availability}
The data that support the findings of this study are available from the corresponding author upon reasonable request.


%

\clearpage
\printfigures
\end{document}